\pdfoutput=1
\documentclass[conference]{IEEEtran}
\IEEEoverridecommandlockouts
\usepackage{cite}
\usepackage{amsmath,amssymb,amsfonts}
\usepackage{algorithmic}
\usepackage{graphicx}
\usepackage{textcomp}
\usepackage{xcolor}
\usepackage{array}
\usepackage{booktabs}
\usepackage{longtable}
\usepackage{tabularx}
\usepackage{tabulary}
\usepackage{blindtext, subfig}
\usepackage{dblfloatfix} % fix for bottom-placement of figure
\usepackage{listings}
\usepackage{amsmath}

\def\BibTeX{{\rm B\kern-.05em{\sc i\kern-.025em b}\kern-.08em
    T\kern-.1667em\lower.7ex\hbox{E}\kern-.125emX}}
\begin{document}

\title{Developing a Process in Architecting Microservice Infrastructure with Docker, Kubernetes, and Istio}
\author{\IEEEauthorblockN{Yujing Wang}
\IEEEauthorblockA{\textit{Department of Mechanical and Mechatronics Engineering} \\
\textit{University of Waterloo}\\
Waterloo, ON, Canada \\
yj9wang@edu.uwaterloo.ca}
\and
\IEEEauthorblockN{Darrel Ma}
\IEEEauthorblockA{\textit{Cheriton School of Computer Science} \\
\textit{University of Waterloo}\\
Waterloo, ON, Canada \\
darrel.ma@edu.uwaterloo.ca}
}

\maketitle

\begin{abstract}
    As an application usage grows, its owner scales up vertically by replacing old machines with more powerful ones. This methodology is expensive and leads to resource waste. In response to the business needs, internet giants have developed the microservice architecture, which lets developers divide up their application into smaller units that can be hosted on multiple machines, thus enabling horizontal scale up. We propose a triphasic incremental process to transform a traditional application into a microservice application that guarantees stability during the operation. Then we demonstrated such methodology in a prototype microservice application based on an existing monolithic application. First, the developer splits a monolithic application into atomic services and aggregated services. Second, these services are packaged, containerized, and then deployed on Kubernetes. During this stage, Istio is deployed on the Kubernetes cluster to establish pod level communications, delegate traffic flows and filter requests, and enable the autoscaler. Other external add-ons, such as database connections, are defined in service entry. In the last stage, we developed an algorithm guideline to minimize inter-service calls by compiling all needed calls into a list and perform one finalized call. Although it increases memory usage, it avoided the wait time incurred during interservice calls. We then investigated managing configurations using config maps, recommended a pipeline being developed to perform automatic rollover. 
\end{abstract}

\section{Introduction}
When an application usage grows, its owner scales it up to handle the increased traffic. Traditionally, companies scale up vertically by replacing current servers with more powerful ones\cite{b1}. This practice requires looking for more voluminous hardware resources at the time of needs, doesn't account for sudden traffic, and requires a major upfront capital investment. When the application does not use machines at their full capacity, resources are wasted. To compensate for the rigidity of vertical scaling, internet giants are promoting the microservice architecture that sees applications decoupled into logical units and then sliced into microservices. These services are packaged in Docker images and deployed on Kubernetes, which handles the hosting, scaling, and monitoring. Then aggregated services are used to facilitate inter-service communications \cite{b2}.

Kubernetes is an open source orchestration system offered by Google for managing containerized services and facilitating declarative configurations and automation\cite{b2}. Although Pivotal offers cloud solution for Java Spring applications\cite{3}, adapting the microservice to a codebase specific platform would create a dependency on Java making it hard to move away should the team rewrites the program in other languages. In addition, adapting to codebase specific platform would be a bad example for developers that seeks a consistent guideline across all codebase. There is an urgent need for a replicable, scalable, and easy to follow process to transform monolithic applications to microservices. This process must be codebase agnostic host agnostic. This paper focuses on developing a methodology to transform traditional Java Spring monolithic backend applications to a network of microservice applications containerized with docker and hosted on Kubernetes. The report will dive deep into the findings during our research progress and discuss concern arose in the process of development.

\subsection{Scaling}
\begin{figure}[htbp]
    \centerline{\includegraphics[width=\linewidth,scale=0.5]{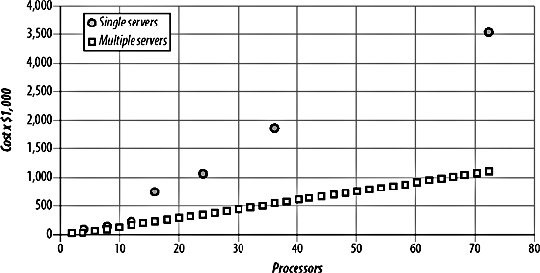}}
    \caption{Cost for Vertical vs Horizontal Scaling \cite{b4}}
    \label{figure1}
\end{figure}

To elaborate on the previously introduced technical terms, vertical scaling and horizontal scaling are not exclusive ideas. Indeed, they are expansions in two different dimensions\cite{b1}. Horizontal scaling requires adding more hardware resources; while, vertical scaling requires finding hardware resources more powerful than current ones. Expanding on both directions yields maximum capacity. However, to attain the most cost-effective way requires finding a balance in the two directions. As Fig.~\ref{figure1} shows, the cost of using less than 15 processors in the server is cheaper than purchasing multiple servers. Having a single server would benefit from using less housing space. Horizontal scale-up adds cost to the storage space and machine maintenance; meanwhile, the cost of adding processors to a single machine grows exponentially. 
\subsection{Hosting Strategy}
The aforementioned "On-premise" is one of the hosting strategies that clients host applications on their own infrastructure, on-premise \cite{b5}. It gives clients total control of their data but requires them to maintain network and security. "On cloud" is where the clients host their application on one of the providers' cloud networks such as Google Kubernetes Engine. Some cloud provider provides private cloud (IBM Bluemix), where the client's data is being put separately on machines on the cloud. 

No upfront cost and flexible scaling make "On cloud" the most appealing option among startups that seek rapid growth and are willing to outsource their network security management to big firms. Compared to on-premise, cloud reduces the initial set up time and removes the need to organize a team of network engineers and security engineers. However, as usage grows, the cost of the cloud also increases exponentially. Snapchat, an instant picture chat startup that uses Google's Cloud Engine, pays Google 2 billion and Amazon 1 billion as of 2018 in cloud contracts \cite{b6}.

\section{Problem Definition}
While we see a surge in demand for a microservices strategy, there lacks a concrete replicable yet scalable methodology to transform monolithic Java application to microservices application. The existing methodology does not offer a clear definition of minimum effort service discovery. Ones offered by Google's guidelines are vague for adapting Java Spring application. We seek to apply the emerging microservice architecture into production for efficient resource management and satisfy increasing client demands for a microservice strategy. We have decided to put the emphasis on platform agnosticism and codebase versatility. There is yet to be developed a portable, scalable, and continually deployable methodology to efficiently transform existing monolithic application to a network of microservices. Specifically, three questions need to be addressed:
\begin{enumerate}
    \item How to perform services discovery and inter-service communication?
    \item What data strategy would guarantee an efficient read and write capability? How to make such a data strategy compatible with existing data infrastructure?
    \item How do we ensure any problems that arose in the operation of the microservice application are picked up in a speedy manner and sent quickly to the right team for a bug fix?
\end{enumerate}
The transformed microservices must be able to perform functions that the original service does. The methodology should make the application more portable, more continually deployable, and more scalable than its original service. By looking at 12 factors app \cite{b7} as criteria, we will place emphasis on 
\begin{enumerate}
    \item service discovery
    \item data strategy
    \item canary development
\end{enumerate}

\section{Breaking up the Monolith}
The first step is to break up an existing application into microservices. In an ideal microservices ecosystem, each service represents what a business unit does \cite{b8}. It exposes an API that the developer uses to communicate with other services. Each microservice should be stateless, meaning an enclosed lifecycle independent from the state of other services. Finally, each microservice enforces a team to program and maintain such a service independently. For demonstration purposes, a mock social media application "Userapp" is developed. It lets the user set up an account, add friends, and make posts. The structure of the java app is shown in Snippet~\ref{code1}.
\begin{lstlisting}[frame=single,caption={Monolithic Java Application Structure}, label={code1} ]
userapp
  java
    src
      Application.java
      controller
        Controller.java
        RestController.java
      service
        Userapp.java
        AccountService.java
        FriendService.java
      Domain
        Account.java
        Friend.java
        FriendType.java
        Message.java
        Post.java
      repositories
        AccountRepositories.java
        FriendRepositories.java
      test
        ControllerTest.java
...
\end{lstlisting}
In which, the service controls the business logic, the userapp.java is where interactions with the submodules are defined as shown in the 3 submodules shown in Table~\ref{table1}: 
\begin{table}[htbp]
\caption{Microservices Broken down from Monolith App}
\begin{center}
\begin{tabular}{|c|c|c|}
\hline
Userapp.Java & AccountService.java & FriendService.java  \\
\hline
{\begin{lstlisting}[frame=none]
Userapp: 
getAccount()
updateAccount()
post()
makeFriend()
\end{lstlisting}} &
{\begin{lstlisting}[frame=none]
Userapp: 
getAccount()
updateAccount()
post()
makeFriend()
\end{lstlisting}} &
{\begin{lstlisting}[frame=none]
Userapp: 
getAccount()
updateAccount()
post()
makeFriend()
\end{lstlisting}} \\
\hline
\end{tabular}\label{table1}\end{center}\end{table}
When the user makes a post, the post function in 'UserApp' will be called to generate the post, which then calls 'AccountService' that register the post onto one's own wall. Subsequently, it calls getAccount() to get a list of the user’s friends from FriendService to notify them that a post has been made by the said user. The repository module handle creates, read, update and delete (CRUD) operations.

\begin{figure}[htbp]
    \centerline{\includegraphics[width=\linewidth,scale=0.5]{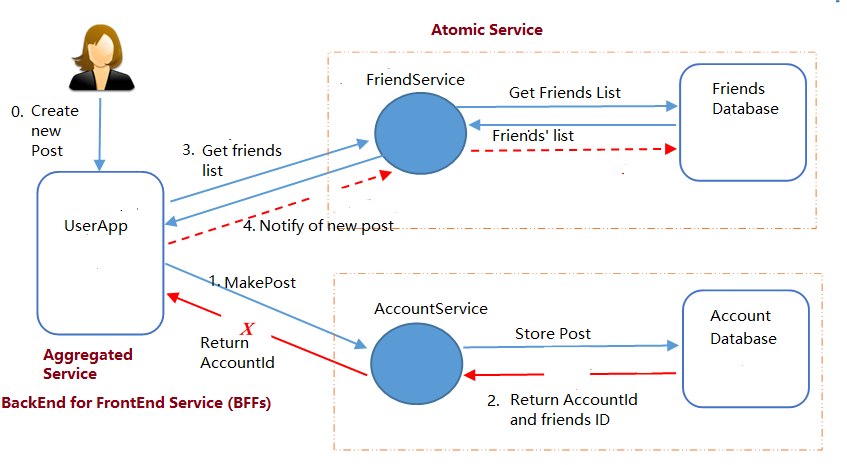}}
    \caption{Architecture View of "Userapp"}
    \label{fig_userapp}
\end{figure}

Three classes in service categories are each responsible for one domain of functions and can be easily be separated into three services. Once the application has been broken down to this level, it cannot be further broken down because functions in each class are tightly coupled and inter-reliant. FriendService and AccountService cannot be further broken down because their functions belong to one set of logic tied to one database. Hence this kind of service is called atomic services. Meanwhile, Userapp does not have any resources attached but works by accepting and sending requests from other services. This kind of service is called an aggregated service, as Fig.~\ref{fig_userapp} shows. In this specific case, userapp service interacts directly with the front end through API and hence is also an example of Backend for Frontend Service (BFFs).

\section{A Triaphasic Incremental Approach}
\begin{figure}[htbp]
    \centerline{\includegraphics[width=\linewidth,scale=0.5]{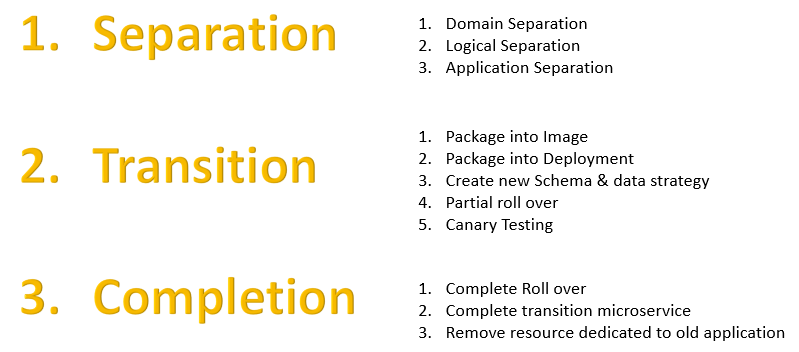}}
    \caption{Triphasic Approach to Transform app. to Microservice}
    \label{fig_triphase}
\end{figure}

Service built from the early days of Object-Oriented Programming (OOD) may not have such a clear cut of separation of concerns unlike 'Userapp'. For example, older applications may have placed friends and accounts in one database. As a result, we realized that the clear-cut idealistic approach is rather impractical, decides to take a progressive transformation in Fig.~\ref{fig_triphase}

\subsection{Seperation}
Using this approach, the development team will first look at the 'domain' category in Table~\ref{table1}. Then identify each domain as primary or helper object, which is shown in Snippet~\ref{helperobj}. A primary object has a set of functions to perform business logic. A helper object does not have its own functions. For example, 'Post' is a helper object that has no function of its own but is used by other functions to fulfill their business logic. 

\begin{lstlisting}[caption={Helper Object Used by Both AccountService and FriendService},label={helperobj},frame=single]
account.makePost(post);
friends.notify(posts);
\end{lstlisting}

Next, A hard logical separation must be placed between these primary objects. In a monolithic application, the account may directly call a friend's database. The developer must remove these cross-domain calls first. Ideally, services on the same hierarchy will never call each other. To bridge the communication, an aggregated service, in this case, Userapp, is used to bridge communication between the services. Because of this separation, some calls that could have been made directly now require exiting the originating service, finding the destination, and then entering the new services. As a result, communication now requires more calls and longer trips. This is a tradeoff that programmers need to make. Generally, developers shall aim for complete separation to avoid having to go back and reconsider logical separation. Once these logical units are packaged into microservices, the perks of microservices such as autonomous scaling, rapid development, and efficient service to service communication can compensate some efficiency lost in inter-service communication. Most importantly, separating the logical units also allows function-specific development teams to develop independently without reliance on other services. Bugs in the service can be quickly allocated and fixed.

\subsection{Transition}
At phase two, "transition", the development team needs to work with the operation team to package the deployment, create new schemas for new atomic services. Each service corresponds to one schema, one database. The database is usually hosted separately from service because services are stateless, but databases are stateful. Next, the developer points the new microservices to the database and labels it as a test environment, running alongside the production environment. As the system matures, the group can decide whether to move a certain percentage of traffic into the new environment. In a progressive approach, this step will progress slowly to ensure minimum disturbance by having the existing monolithic application running alongside the new microservice app. Ultimately, this is a tradeoff between stability and resource. In the last phrase, the team will move everything to the new application and remove reference to the old. By now, the microservice application is mature, and the operation team will take over the app.

Breaking up the Monolith is the first step in transforming a monolith application into a microservice. A progressive approach prevents having to go back to re-engineer the service. Small changes are made incrementally towards building a big system. Developers should keep in mind the tradeoff they make when separating logical units and have reasonable logic before making each decision. The lower level the blocks split, the higher the cost to each call across the individual unit, but also, the easier it is to convert into microservice and scale-up. For large scale applications, it is always better to perform complete separation and have the infrastructure to handle the scaling for better performance. Relate back to the criteria and constraints, properly laying out the foundations for microservice allows for a portable and scalable and continually deployable methodology.

\section{Building Deployments}
Packaging the sliced application from phrase one into deployments is a necessary step in phrase two. At this stage, both Docker and Kubernetes will be used to create deployment \cite{b2}. The deployer will only need the packaged image to complete this task. Ad-hoc modifications such as changing the port name or hard coding a destination in source code to aid deployment are strictly prohibited. Otherwise, changes made by one developer would result in the methodology being inadaptable to other applications.

\subsection{Docker Deployment}
Docker lets the developer run applications in any environment that has a Docker engine without having to worry about dependencies and the OS environment \cite{b9}. The user writes a Dockerfile, as shown in Snippet 4, to build the Docker image. The developer specifies the source code and operating system in the "FROM" line separated by ":". Then copies the original executable file using the "COPY" command into the image. Other relevant commands can be executed with "CMD". Lastly, expose a port for API communication with \{application\_port\}:\{Image\_port\} command found at the last line of Snippet~\ref{javadockerfile}. 
\begin{lstlisting}[frame=single,label={javadockerfile}, caption= {Dockerfile to build a Java Spring app. Image}]
FROM openjdk:8-jre-alpine
COPY userapp-0.0.1-SNAPSHOT.jar / app.jar
# run application with this command line
CMD ["/usr/bin/java", "-jar", "-Dspring.profiles.active=container", "/app.war"]
EXPOSE 8080:8080
\end{lstlisting}

Then run the Dockerfile with Snippet~\ref{dockerbuild} in the terminal to build the docker image. In this statement, the author supplies an image tag (userapp) and version (latest). The tag identifies an image registry where the image will be stored. If no version has been supplied, the "latest" version will be the default.
\begin{lstlisting}[frame=single, label={dockerbuild}, caption={Command to Build a Docker Image with a Tag}, language=bash]
$ docker build -t userapp:latest .
\end{lstlisting}

\subsection{Kubernetes Deployment}
\begin{figure}[htbp]
    \centerline{\includegraphics[width=\linewidth,scale=0.5]{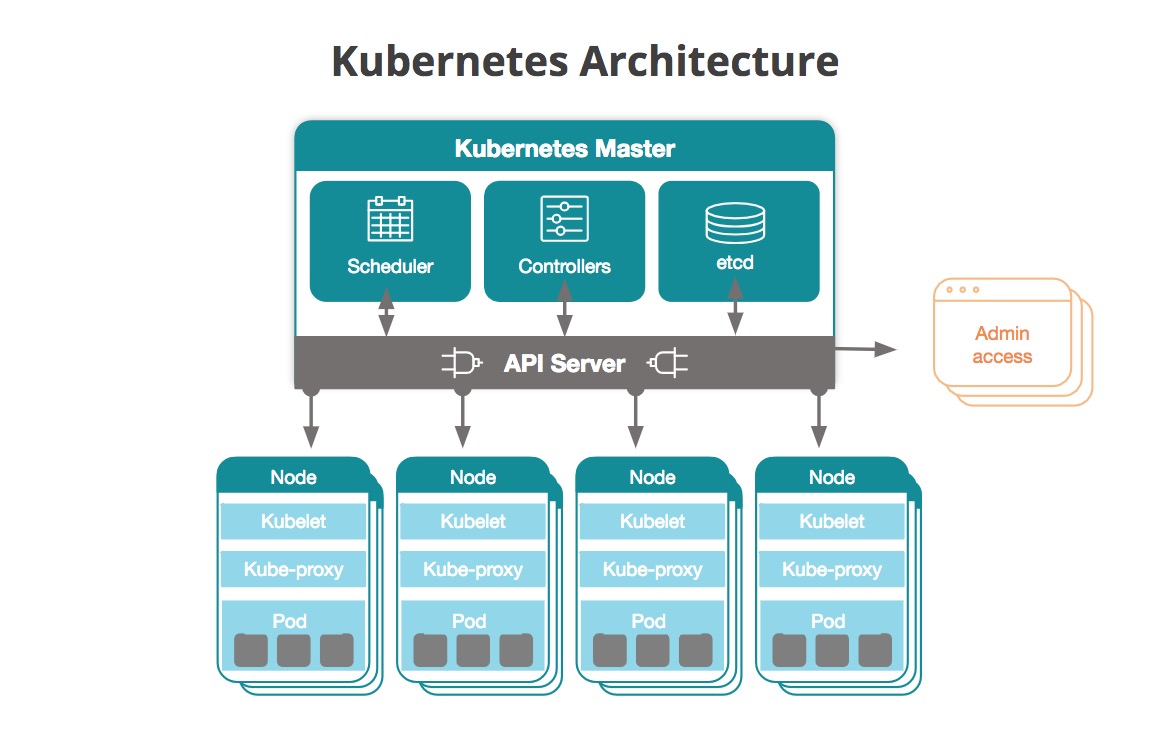}}
    \caption{Arcitecture of Kubernetes}
    \label{fig_kubernetes_arch}
\end{figure}

Kubernetes creates deployments from Docker images, establish service communication with envoy gateways that looks up services' IP address\cite{b10}. When a service's usage increases to a preset threshold, Kubernetes Autoscaler replicates the service to fulfill the increased load. Kubernetes follows a master-workers architecture Fig.~\ref{fig_kubernetes_arch}. The master node accepts commands, controls cluster, schedules pods, and store configurations using kube-apiserver, kube-controller, kube-scheduler, and etcd. Then it uses Kubelet to perform the action on a node level. When the user executes Snippet~\ref{kubernetesdeplpyment} using kubectl.
\begin{lstlisting}[frame=single,label={kubernetesdeplpyment}, caption={Command to Create Kubernetes Deployment}]
$ kubectl create -f deployment -s 8080
\end{lstlisting}
Kubernetes' API server receives the request from keyword 'kubectl' and delegates the controller manager to create a new deployment from the 'deployment' file. After creation, the scheduler takes over the newly created pod on an available thread and expose it to port '8080' \cite{b2}. The most fundamental building block of Kubernetes is a pod. Similar to the idea of the container from docker, pods are containerized units in Kubernetes with a specific IP address assigned by Kubernetes \cite{b2}. A pod can be created, cloned, terminated, and destroyed by the node. As shown in Figure 6, a pod may containerize an app, a volume, or both.

To customize a kubernetes deployment, the developer writes an ymal configuration according to Snippet~\ref{kub_deploy} \cite{b2}:
\begin{lstlisting}[frame=single, caption={Kubernetes Deployment}, label={kub_deploy}]
controllers/userapp.yaml 
apiVersion: apps/v1
kind: Deployment
metadata:
  name: userapp-deployment
  labels:
    app: userapp
spec:
  replicas: 3
  selector:
    matchLabels:
      app: userapp
  template:
    metadata:
      labels:
        app: userapp
    spec:
      containers:
      - name: nginx
        image: userapp:latest
        ports:
        - containerPort: 8080
\end{lstlisting}
The first line defines the Kubernetes API version. 'Kind' in the second line specifies the type of service this configuration contains. The "kind" field expects the type of Kubernetes object, the scheduler will expect. The most used "kinds" are pod, deployment, service, and PV. As mentioned before, 'pod' is a group of containers, including storage units that made up the most fundamental building block of Kubernetes \cite{b2}. When a pod is deployed, it only exists as a single instance and can be communicated through the kubectl API. It is usually used to host testing tools inside the Kubernetes container. For example, when the developer needs to test autoscaling, it becomes unrealistic to send 5000 requests in one second manually. A pod with 'busybox' image is created. Busy boy would continuously spam requests to the service at a specified interval with specified volume.

\begin{figure}[htbp]
    \centerline{\includegraphics[width=\linewidth,scale=0.5]{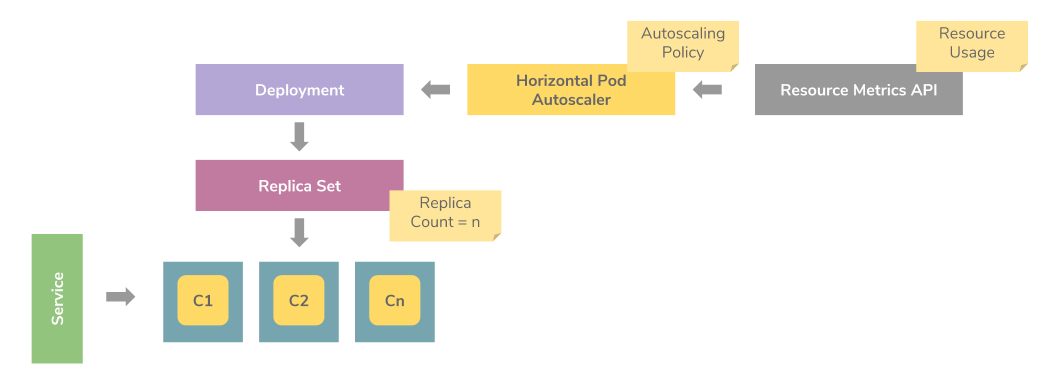}}
    \caption{Kubernetes Deployment Flow}
    \label{fig_kubedeployflow}
\end{figure}

Very similar to the pod, "deployment" is another Kubernetes'  kind' \cite{b2}. When creating a deployment, Kubernetes instantiates one pod with one replica set. If more traffic is being routed to this service, deployment can replicate its pods according to its replica set. Deployment lets the user control all instances created from one deployment at once. That way, an ordinary Kubernetes developer would not need to manually go through countless pods to apply the same change to one application. Detailed workflow of $deployment->replica set->pods$ are shown in Fig.~\ref{fig_kubedeployflow}.

Kubernetes offers more fine-grained resource control capability on deployment level than on the pod level. Take checking the status of pods as an example. Simply run "get pod" command in Snippet~\ref{getpod}, the developer will find three pods belong to the same deployment, yet each pod has a different id. 
\begin{lstlisting}[frame=single, label={getpod}, caption={Get Pods with Label}]
$ kubectl get pods --show-labels
NAME READY   STATUS  RESTARTS AGE
nginx-deployment-75675f5897-7ci7o       1/1     Running     0 18s
nginx-deployment-75675f5897-kzszj       1/1     Running    0  18s
nginx-deployment-75675f5897-qqcnn       1/1    Running    0 18s
\end{lstlisting}
To update the deployment, the developer executes Snippet~\ref{delandcreate} as many times as there are pods.
\begin{lstlisting}[frame=single,label={delandcreate}, caption={Manually Updating Pods One by One}]
kubectl delete pod nginx-deployment-{deployment_id}-{pod_id}
kubectl create pod {name_pod} -f {name_newapp}
\end{lstlisting}
However, the developer can work directly with deployment. First, execute Snippet~\ref{kubgetdeploys} to check deployments and find that there are three pods associated with the “nginx-deployment”.
\begin{lstlisting}[frame=single,label={kubgetdeploys},caption={Get Deployments}]
$ kubectl get deployments
NAME               DESIRED   CURRENT   UP-TO-DATE   AVAILABLE   AGE
nginx-deployment   3         3         3            3           18s
\end{lstlisting}

The developer runs Snippet~\ref{editdeploy} to perform an update:
\begin{lstlisting}[frame=single,label={editdeploy}, caption={Edit Deployments with New Configurations}]
$ kubectl edit deployment.v1.apps/nginx-deployment
deployment.apps/nginx-deployment edited $
\end{lstlisting}
In order to see what has happened, the user runs the describe command Snippet~\ref{describedeploy}.
\begin{lstlisting}[frame=single, label={describedeploy}, caption={Describe Deployments to see Events and Logs}]
$ kubectl describe deployments
Name:                   nginx-deployment
Namespace:              default
CreationTimestamp:      Thu, 30 Nov 2017 10:56:25 +0000
Labels:                 app=nginx
Annotations:            deployment.kubernetes.io/revision=2
Selector:               app=nginx
Replicas:               3 desired | 3 updated | 3 total | 3 available | 0 unavailable
StrategyType:           RollingUpdate
MinReadySeconds:        0
RollingUpdateStrategy:  25% max unavailable, 25% max surge
Pod Template:
  Labels:  app=nginx
  Containers:
   nginx:
    Image:        nginx:1.9.1
    Port:         80/TCP
    Environment:  <none>
    Mounts:       <none>
  Volumes:        <none>
Conditions:
  Type           Status  Reason
  ----           ------  ------
  Available      True    MinimumReplicasAvailable
  Progressing    True    NewReplicaSetAvailable
OldReplicaSets:  <none>
NewReplicaSet:   nginx-deployment-1564180365 (3/3 replicas created)
Events:
  Type    Reason             Age   From                   Message
  ----    ------             ----  ----                   -------
  Normal  ScalingReplicaSet  2m    deployment-controller  Scaled up replica set nginx-deployment-2035384211 to 3
  Normal  ScalingReplicaSet  24s   deployment-controller  Scaled up replica set nginx-deployment-1564180365 to 1
  Normal  ScalingReplicaSet  22s   deployment-controller  Scaled down replica set nginx-deployment-2035384211 to 2
  Normal  ScalingReplicaSet  22s   deployment-controller  Scaled up replica set nginx-deployment-1564180365 to 2
  Normal  ScalingReplicaSet  19s   deployment-controller  Scaled down replica set nginx-deployment-2035384211 to 1
  Normal  ScalingReplicaSet  19s   deployment-controller  Scaled up replica set nginx-deployment-1564180365 to 3
  Normal  ScalingReplicaSet  14s   deployment-controller  Scaled down replica set nginx-deployment-2035384211 to 0
\end{lstlisting}

The developer immediately noticed the replica set section and found out that three pods are running, and all of them have been successfully updated. The exact step of an update is found in the 'Event' section. When the deployment was created initially, three pods have been created. Since the three pods were created from nothing, this action was classified as scale up. Then the user runs the update command; the second event fired up to create one copy of the new deployment, another scale-up. Kubernetes detected that four pods are now running and are therefore above the maximum threshold, it fires up the 3 rd event: a scale down from 3 to 2 for the original pod. This cycle runs until all the original pod is shut down and three new pods up and running.

This cycle that Kubernetes utilize is called "rollover" \cite{b2}. A strategy commonly practiced by network platform engineers to ensure stability during the update, meanwhile minimizing memory resource in an update. If there are N pods and all need to be updated, the maximum memory occupied is N+1 pod. And since the original pods already occupy N spaces. (N+1) – N = 1 space is used, hence O(1), constant memory, which is the lowest memory usage achievable. This method also guarantees that at any time, there is at least N pod running. If three people are using the application during the update, one person would soon notice the change, then the second, then the third. None of them would be forced offline.

After creating the deployment, the developer creates service to expose deployments to the communication channel by executing Snippet \ref{servicecreate}.

\begin{lstlisting}[frame=single,label={servicecreate}, caption={Creating Service on Kubernetes}]
apiVersion: v1
kind: Service
metadata:
  name: nginx
  labels:
    run: nginx
spec:
  type: NodePort
  ports:
  - port: 8080
    targetPort: 80
    protocol: TCP
    name: http
  - port: 443
    protocol: TCP
    name: https
  selector:
    run: nginx
\end{lstlisting}

\begin{figure}[htbp]
    \centerline{\includegraphics[scale=0.8]{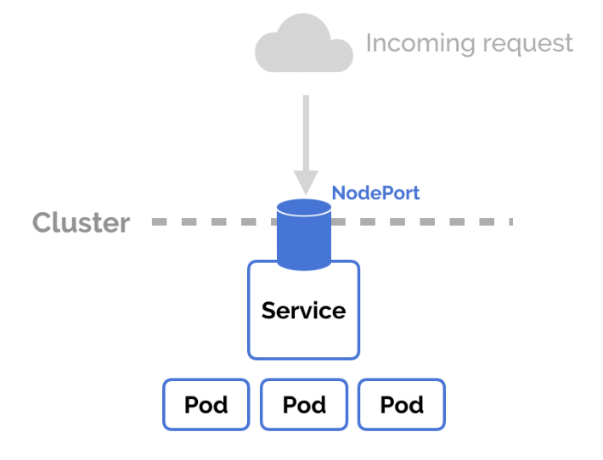}}
    \caption{Nodeport}
    \label{fig_node}
\end{figure}

For each port definition, the user will provide an originating 'port' that listen for request and a 'targetPort' on deployments that the service forward to, and a connection protocol. Similar to how people make calls to www.google.com instead of https://www.google.com:80, in Kubernetes, the default protocol TCP and default port 80 can be left out. Note at spec.type, 'nodeport' is specified. This specifies the type of service. Kubernetes' default service type is Cluster-IP, which is only available within the cluster. Node Port exposes the service' node and port, as seen in Fig.~\ref{fig_node}.

\begin{figure}[htbp]
    \centerline{\includegraphics[width=\linewidth,scale=0.5]{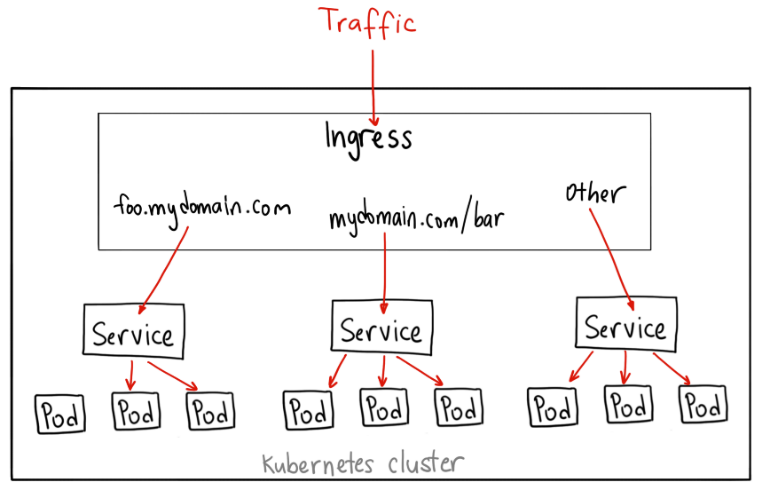}}
    \caption{An Ingress Gateway}
    \label{fig_ingress}
\end{figure}

\begin{figure}[htbp]
    \centerline{\includegraphics[width=\linewidth,scale=0.5]{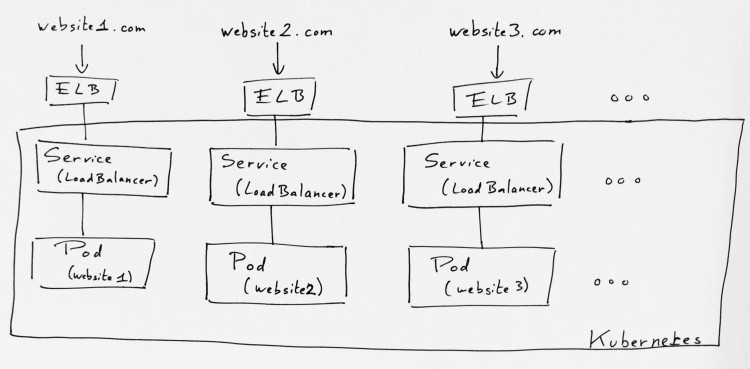}}
    \caption{Elastic Load Balancer}
    \label{elas}
\end{figure}

The developer may directly access the service from their computers without ssh into the Kubernetes cluster. Nodeport exposes a vulnerability in the system that is prone to a Distributed Denial of Service (DDoS) attack, where the perpetrator uses a distributed system to spam requests to the node port system result in legitimate service lost in a flood of bot requests. Although some may argue that one can use nodeport during development and, subsequently, remove such vulnerability in production. This still cultivates a reliance on Nodeport for developers, which leads to a bad habit hard to change. This clear-cut solution removes human developers from developing such a tendency. The team has decided to ban the use of node port such that all traffic will be forced to go through a gateway coupled with a security layer, as Fig.~\ref{fig_ingress} shows.
Another explicit way to expose service is 'Load Balancer'; it exposes the port to an external load balancer. Typically, this is used by services hosted on a cloud platform, which has provided a load balancing function, and the client need not perform load balancing task. Fig.~\ref{elas} shows the architecture of one that hosts Kubernetes on Amazon Web Service. The web app supports entry from website1.com, website2.com, and website3.com through Amazon's Elastic Load Balancing (ELB) into service defined as "load balancer" service \cite{b11}. Since the Kubernetes is hosted on the cloud, traffic coming from ELB has already passed through Amazon's security layer.
Clearly, this situation does not apply to the team that is developing a solution for hosting on-premise. A special type of service is called "virtual service" provided by Istio, a service mash manager developed by Google \cite{b12}. Virtual service, coupled with destination rules, defines a set of traffic routing rules to apply when a host is addressed. Each routing rule defines matching criteria for the traffic of a specific protocol. If the traffic is matched, then it is sent to a named destination service defined in the registry.

The last "kind" is Persistent Volume(PV). PV can be set up within a pod under the "spec" section or as its own pod \cite{b2}. Kubernetes uses volume to store data. To create volume, initiate a Persistent Volume Claim (PVC), a request for a PV. PVC defines the read and write authority of incoming traffic, and manage resources consumed by PV. Then the developer creates a PV that spawns up a stateful pod within the cluster where data is stored. A PV must exist corresponding to a PVC if the developer attempts to access the data. Otherwise, the PVC will dynamically spawn up PV pods to store the data. This method is used when Kubernetes wants to store some temporary data and only need the system to access it. However, the team discovered that storing stateful data in stateless pods and by labeling them stateful is rather paradoxical. The system would still treat these labeled stateful pods as stateless and shut down pods from time to time, resulting in data loss.

\begin{figure}[htbp]
    \centerline{\includegraphics[width=\linewidth,scale=0.5]{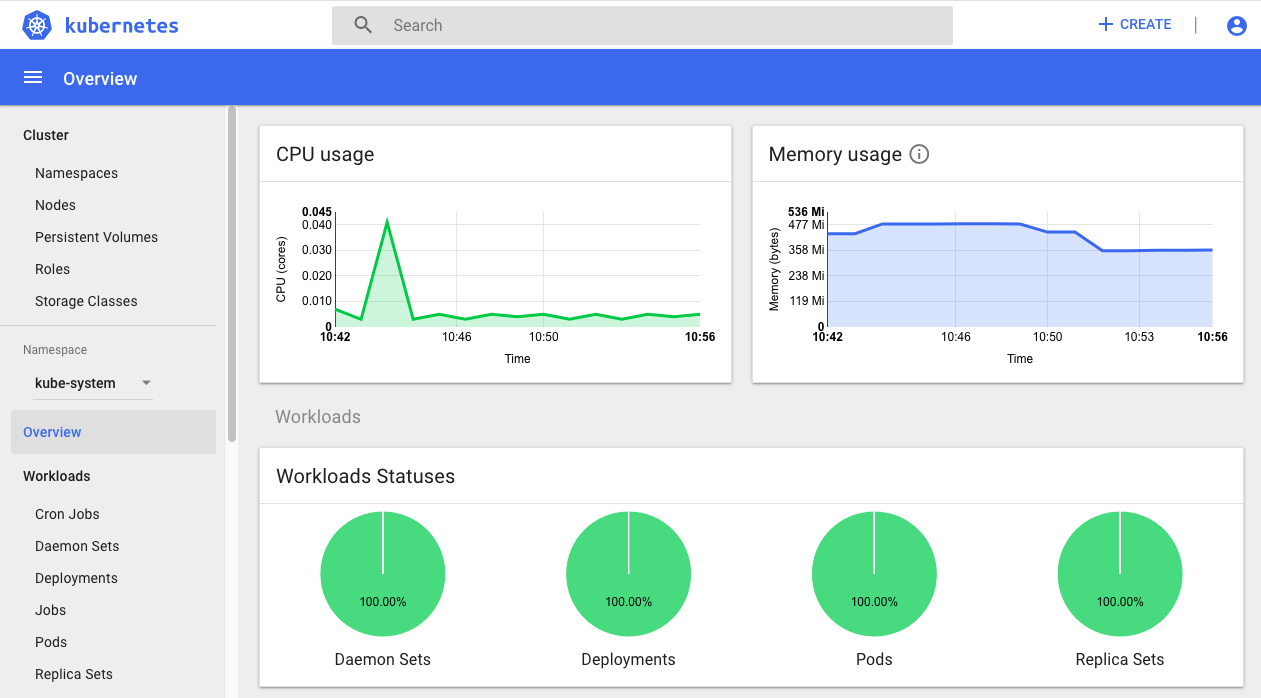}}
    \caption{Dashboard for Kubernetes}
    \label{fig_dashboard}
\end{figure}

Moreover, spawning a persistent storage every time when the system initiates takes time. In Figure 13, only one pod containing a 'busybox' has been initiated and running, and it costs 400 MiB and at the peak of CPU usage 0.045 cores, as shown in Fig.~\ref{fig_dashboard}. Thus, PV has not been investigated in detail for this project.

\section{Service Mesh}
\begin{figure}[h]
    \centerline{\includegraphics[width=\linewidth,scale=0.5]{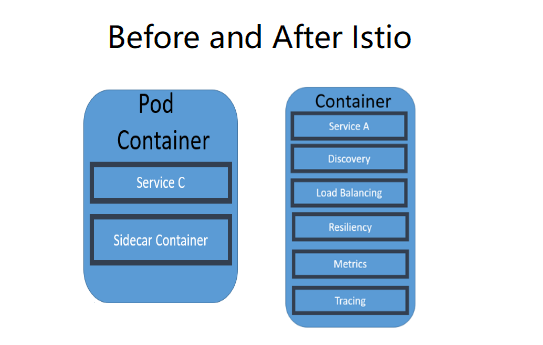}}
    \caption{Istio Sidecar Injection}
    \label{istio}
\end{figure}

\begin{figure}[htbp]
    \centerline{\includegraphics[width=\linewidth,scale=0.5]{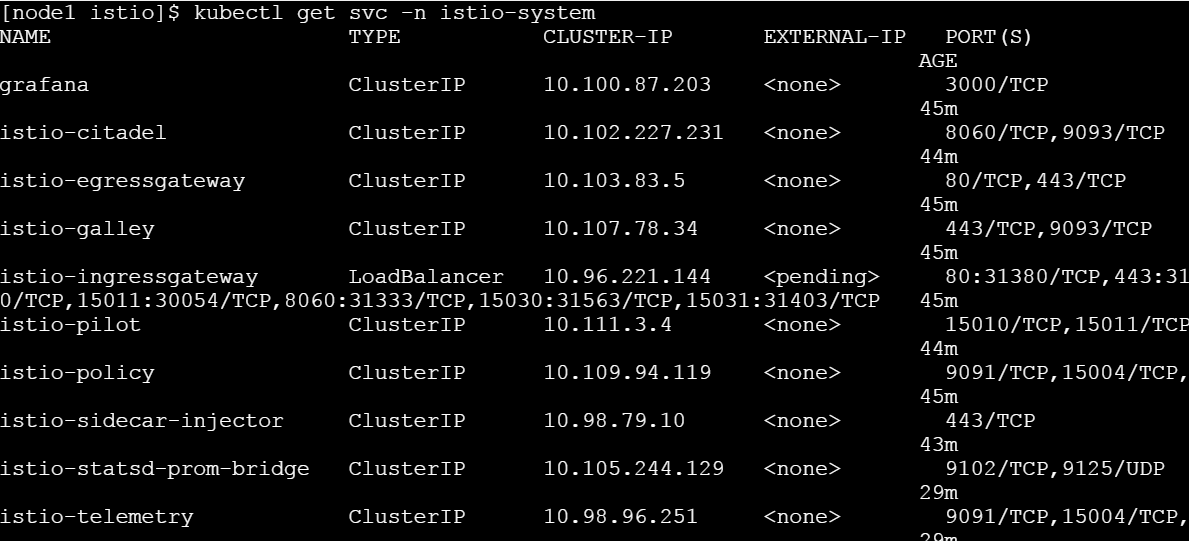}}
    \caption{Checking Istio Installation Services}
    \label{istiocheck1}
\end{figure}

\begin{figure}[htbp]
    \centerline{\includegraphics[width=\linewidth,scale=0.5]{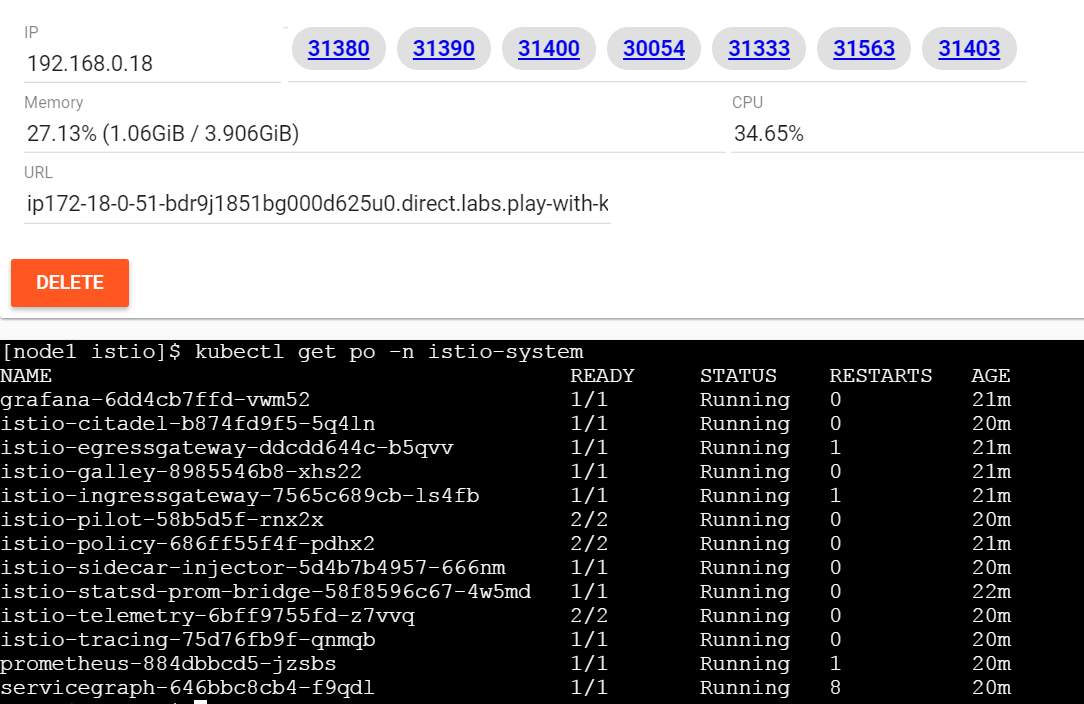}}
    \caption{Checking Istio Installation Pods}
    \label{istiocheck2}
\end{figure}

A service mesh is a dedicated infrastructure layer to run a fast, reliable, and secure network of microservices, a container orchestration system to provide a high level of deployment infrastructure \cite{b12}. Although the Kubernetes system provides the bare minimum backbone for service to service communication, it is far from easy to work with in terms of supporting a robust planetary-scale application. A couple of issues needs to be addressed:
\begin{enumerate}
    \item How to communicate between services effectively
    \item How to handle load balancing in the mesh without external Load Balancer
    \item How to monitor each
\end{enumerate}
In order to handle these issues, the team lead has investigated several solutions and decided to use Istio for its high degree of customization. The team is tasked with looking into extracting the useful aspect of Istio and applying them to the microservice prototype. Fig.~\ref{istio} shows how Istio works by injecting sidecar into pods to perform mesh commands from Istio's master node.
To install Istio, run script Snippet \ref{istioinstall}:
\begin{lstlisting}[caption={Istio Installation},label={istioinstall}]
#!/bin/bash
curl -L https://git.io/getLatestIstio  | (sh -)
export PATH=$PWD/bin:$PATH
cd istio-1.0.0
kubectl apply -f install/kubernetes/helm/istio/templates/crds.yaml
kubectl apply -f install/kubernetes/istio-demo.yaml
kubectl get svc -n istio-system
kubectl get pods -n istio-system
\end{lstlisting}
The developer then runs Snippet \ref{istiocheck1} to check if service has been installed and Snippet \ref{istiocheck2} to check deployment.

\begin{figure}[htbp]
    \centerline{\includegraphics[width=\linewidth,scale=0.5]{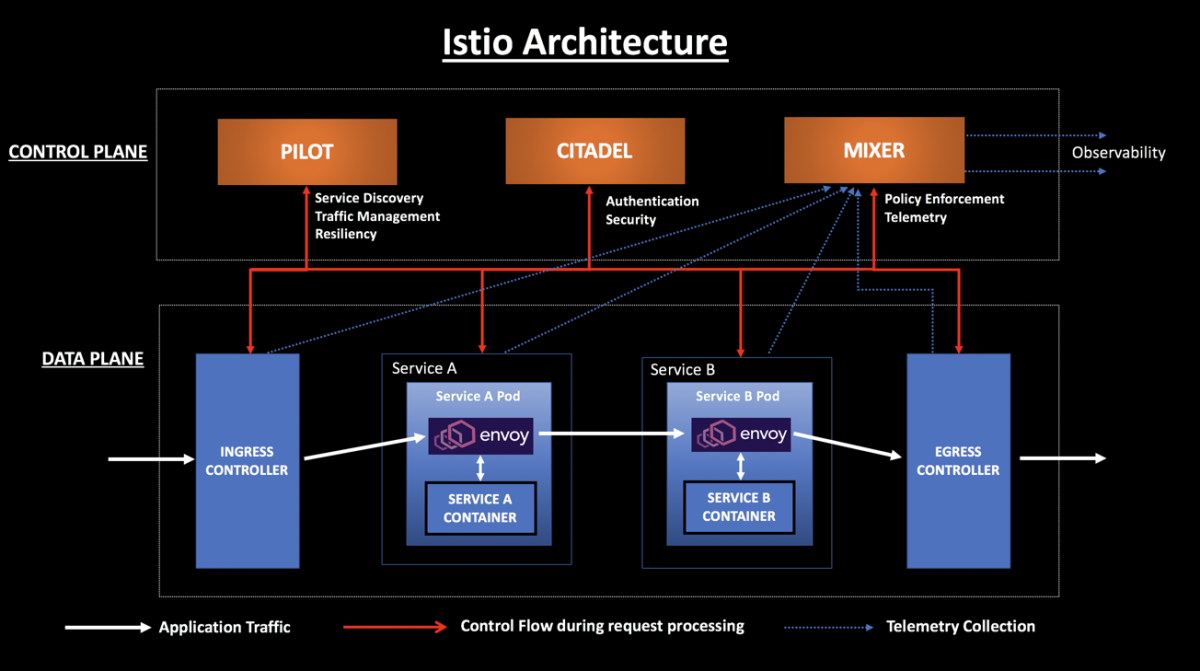}}
    \caption{Istio Architecture}
    \label{istioarch}
\end{figure}

If all pods are "ready" and "running", Istio has been deployed successfully. After deployment, Istio, together with Kubernetes, will consume approximately 1 GiB of memory as Snippet 14 shows. The memory consumption will increase logarithmically as more pods being instantiated. Istio works by injecting a sidecar into the pods, as shown previously in Fig. \ref{istio}. The developer just needs to label the namespace with sidecar injection enabled by executing Snippet \ref{enableinject}, Enabling Sidecar Injection on 'default' namespace. 
\begin{lstlisting}[label={enableinject}, caption={Enable Istio Injection}]
$ kubectl label namespace default istio-injection=enabled
\end{lstlisting}
The developer now can use Istio and all its functionality. Istio is made of two components (Fig. \ref{istioarch}): a control pane and a data pane \cite{b12}. Much like Kubernetes master workers architecture. The control plane is the engine that offers the user an entry point; meanwhile, the data plane is the components injected into each pod to facilitate Istio functions. The architect of Istio does not want the developers to think of microservice in terms of master and workers. Instead, they what to enforce a parallel development mindset onto developers when developing the microservice architect. The developer may program with deployments and services simultaneously without having first to define all the services and bind them. When the developer spawns up deployment, a service is created by Istio and can be discovered by Envoy, thus removing the need to create service to expose deployments.
Different from Kubernetes' service, at the core of Istio service discovery, is a technology called "Pilot" Pilot manages and configures all the envoy proxy deployed in the sidecar that the developer has previously embedded in each pod. Pilot lets the developer specify rules regarding traffic routing between envoy proxies and define a clear method to handle failure recovery in events such as time out and circuit breaking. Note that the Pilot has embedded load balancing when the traffic travels through the Envoy. Figure 16 shows the working principle of Pilot \cite{b12}. SvcA does not need to have the knowledge of the deployment to access pods in deployment. Previously the service must have defined the access to SvcB Pod1 or svcB Pod2...svcB Pod4 to access the deployment.

\section{Traffic Access}
A few common practices when it comes to defining traffic control are round-robin and weight. Round robin is the default mode, where each request will go to the subsequent instances, hence evenly distribute the traffic. In the weighted traffic access, the user assigns a weight to different deployments, as shown in Snippet \ref{virtualservice}, so the traffic is split according to user-defined weight.
\begin{lstlisting}[label={virtualservice}, caption={Defining Virtual Service to Configure Traffic Split Weight}]
apiVersion: networking.istio.io/v1alpha3
kind: VirtualService
metadata:
  name: reviews
spec:
  hosts:
    - reviews
  http:
  - route:
    - destination:
        host: reviews
        subset: v1
      weight: 75
    - destination:
        host: reviews
        subset: v2
      weight: 25
\end{lstlisting}
The user may also define a timeout and number of retries (Error! Reference source not found.) when the request initially fails to travel through. By industry standard, five retries and 200ms are designated. The time out is calculated based on relaying timeout. That is a gateway to service timeout < 2000ms. Now, let L be a number of layers, and t be timeout defined in milliseconds, and user side time out < 10s, and number of ties (R)>= 1, one for the initial request, and a safety factor of 3, let C be constant time from user machine to Gateway, we uses a system of equation, 
\begin{equation}\begin{cases}t×L<2000ms\\ 3(t×L)+C<2000ms \end{cases}, \end{equation}
with limits obtained by, 
\begin{equation}
\begin{array}{l@{{}={}}l}
    Limits & 3(X<2000) + C<10000 \\ 
    & (3X< 6000)+C<10000 \\
    &  C<10000-(3X<6000) \\
    & C<10000-6000 \\
    & C<4000
\end{array}.
\end{equation}

Hence assume X = $X_{max}$, C=4,000. The machine needs to connect in 4 seconds. However, C can be much lower if the host is a cloud provider. For example, Ping amazon at Toronto gives 6ms, so C= 4 is much smaller than 4000, and X can be much greater. However, for On-premise, we assume the worst-case scenario for C that is the smallest delay at L. Since
\begin{equation}
     c = 4000, 3tl = 6000, tl = 2000, t>0, L>0,  L\in Z
\end{equation}

To obtain L, count the number of layers including platform itself:
\begin{align}
Kubernetes: L0, userapp: L1, account = L2, \\
friends = L2: total = 3, 2000 = t(3), so t = 667  
\end{align}
Apply the answer to write the virtual service in Snippet \ref{ansvirtual}.
\begin{lstlisting}[label={ansvirtual}, caption={Virtual Service with Calculated Timeout Limit}]
apiVersion: networking.istio.io/v1alpha3
kind: VirtualService
metadata:
  name: userapp
spec:
  hosts:
    - userapp
  http:
  - route:
    - destination:
        host: userapp
        subset: v1
    retries:
      attempts: 3
      perTryTimeout: 667ms
    - destination:
        host: account
        subset: v1
    retries:
      attempts: 3
      perTryTimeout: 667ms
    - destination:
        host: friends
        subset: v1
    retries:
      attempts: 3
      perTryTimeout: 667ms
\end{lstlisting}

\section{Data Strategy}
\begin{figure}[htbp]
    \centerline{\includegraphics[width=\linewidth,scale=0.5]{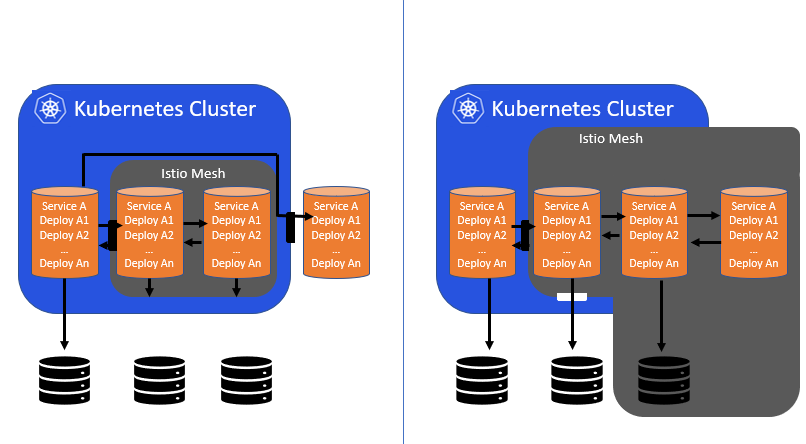}}
    \caption{Mesh Expansion}
    \label{fig_mesh_expansion}
\end{figure}

Istio also offers three ways to handle data strategy \cite{b12}. Mesh expansion is the ideal way. It elevates the items from outside mesh to enjoy the same privilege as if it is inside the mesh. As it shows in Fig. \ref{fig_mesh_expansion}-left, services in mesh(green) cannot access the database outside the Kubernetes. It can access services in Kubernetes but not in Istio through a gateway(black). After a service expansion (Fig. \ref{fig_mesh_expansion}-right) on database and external service, the services can directly access each other. Meanwhile, the service inside Istio can access the database through a service entry(white).
However, as of Istio 1.3.0, Mesh Expansion is not ready according to Google, it only works 50\% of the time. The production environment cannot tolerate such instability. Another choice is to host data storage directly using the built-in Kubernetes capability. That is to utilize PV and PVC, as mentioned in 6.3. Kubernetes lets developers make claims to storage space, then spawn up pods to realize the storage facility. Such an option was rejected as storing data in Kubernetes' pod is basically storing stateful content in stateless infrastructure. Data can be lost when the system decides to shut down pods. As Figure 17 shows our last option: the service can connect to the outside database through an open gate (White) called "service entry". Service entry opens an "entry" into the service and lets only data to be transmitted. Action commands cannot be communicated through service entry; they must go through a gateway. The team decided to use service entry to connect the microservice to the existing database for the time being and leave the schema creation and database break down until Google fixed mesh expansion. 

\section{Canaray Development}
In the "updating pod" example, we demonstrated Kubernetes' capability of rolling out. Kubernetes recursively spawns a new pod then takes down an old one until all old pods are removed \cite{b12}. While Kubernetes' performance is stellar on a deployment level, Istio is needed to handle more robust yet fine-grained rollout on service level. Take the example of updating 1\% of 1000 instances of account service: if a command to update 10 of the pods is ran, the developer does not know which ten pods are updated. To address this problem, the developer can choose to create a new deployment under a different label. However, under a new label, all configurations from the deployment are not inherited. The autoscaler from the old deployment does not control the new deployment. Istio offers control over the rollout process by limiting traffic going into new pods upon initialization, then gradually increases the traffic into the new pods as they mature. If errors happen, Istio rolls back to the previous version. More sophistically, Istio offers control over the region, user, or other properties at the developer's request. 

\begin{table}
  \label{iscompa}  
\begin{tabular}{|c|c|}
\hline plain &  Istio \\ \hline
{\begin{lstlisting}[frame=none]
kind: Deployment
metadata:
  name: helloworld-v1
spec:
  replicas: 1
  template:
    metadata:
      labels:
        app: helloworld
        version: v1
    spec:
      containers:
      - image: helloworld-v1
        ...
---
apiVersion: extensions/v1beta1
kind: Deployment
metadata:
  name: helloworld-v2
spec:
  replicas: 1
  template:
    metadata:
      labels:
        app: helloworld
        version: v2
    spec:
      containers:
      - image: helloworld-v2
        ...
\end{lstlisting}} & 
{\begin{lstlisting}[frame=none]
kind: VirtualService
metadata:
  name: helloworld
spec:
  hosts:
    - helloworld
  http:
  - route:
    - destination:
        host: helloworld
        subset: v1
      weight: 90
    - destination:
        host: helloworld
        subset: v2
      weight: 10
---
apiVersion: networking.istio.io/v1alpha3
kind: DestinationRule
metadata:
  name: helloworld
spec:
  host: helloworld
  subsets:
  - name: v1
    labels:
      version: v1
  - name: v2
    labels:
      version: v2
\end{lstlisting}} \\   \hline
    \end{tabular}

\caption{Plain vs Istio Inject Kubernetes Config}
\end{table}

Table \ref{iscompa} shows how Istio differs from Kubernetes. In plain Kubernetes, the developer needs to create two deployments with different labels, and each needs the specific number of pods specified at "replica". In Istio, the developer creates a new "kind" called virtual service, in which she specified the weight of traffic going into each respective version. Now a simple test can be performed to see the effect. First, run Snippet \ref{autoscaling} to enable autoscaling: 

\begin{lstlisting}[label={autoscaling}, caption={Enable Autoscaler on Kubernetes Deployment}]
$ kubectl autoscale deployment helloworld-v1 --min=1 --max=10
$ kubectl autoscale deployment helloworld-v2 --min=1 --max=10
\end{lstlisting}
Then after spawning a few minutes of request, one will notice that the first deployment scales up much faster than the second one, corresponding to the 9:1 ratio specified in the virtual service (Snippet \ref{afterauto}):
\begin{lstlisting}[label={afterauto}, caption={Check Autoscaler Effect}]
$ kubectl get pods | grep helloworld
helloworld-v1-3523621687-3q5wh   0/2       Pending   0          15m
helloworld-v1-3523621687-73642   2/2       Running   0          11m
helloworld-v1-3523621687-7hs31   2/2       Running   0          19m
helloworld-v1-3523621687-dt7n7   2/2       Running   0          50m
helloworld-v1-3523621687-gdhq9   2/2       Running   0          11m
helloworld-v1-3523621687-jxs4t   0/2       Pending   0          15m
helloworld-v1-3523621687-l8rjn   2/2       Running   0          19m
helloworld-v1-3523621687-wwddw   2/2       Running   0          15m
helloworld-v1-3523621687-xlt26   0/2       Pending   0          19m
helloworld-v2-4095161145-963wt   2/2       Running   0          50m
\end{lstlisting}

To filter request by a specific case, the developer can use the match modifier in Snippet \ref{matchfunction}
\begin{lstlisting}[label={matchfunction}, caption={Uses Match Function}]
http:
  - match:
    - headers:
        cookie:
          regex: "^(.*?;)?(email=[^;]*@company.com)(;.*)?\$"
\end{lstlisting}

Snippet 21 shows the user filters incoming requests containing email ending with domain "@company.com", then delegate all traffic matching this condition to v1 or v2 as shown in Table 3. If the developer wants to delegate all traffic from the United States to V2 and Canada to V1, the developer can have programmed a match for "location", and in the app, but the location in the header whenever the user sends the request. The match will then intercept the location in each request and delegate the traffic to the corresponding destination.

\subsection{Developing an Efficient Aggregated Service Algorithm }
\begin{figure}[htbp]
    \centerline{\includegraphics[width=\linewidth,scale=0.5]{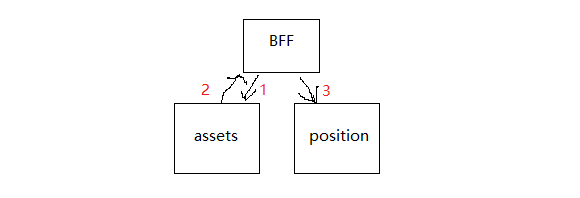}}
    \caption{Sigal Path}
    \label{signalpath}
\end{figure}

During the development of the BFF service, a couple of function challenged the developer's ability to program with microservice. One is notably getting total post count. Traditionally, a SQL script is run in the database that contains both the account and the friends. However, according to the aforementioned logical separation guideline, one should never have more than one domain in one database. The SQL solution is quickly eliminated. The first algorithm developed was Snippet \ref{oldalgo}. The developer does a parallel stream on posts service to get a list of posts the originating user's account. Whilst getting the posts, get posts' like from this persons' friend using the Post ID. The quantity with price and sum all up likes to return total likes. 
\begin{lstlisting}[label={oldalgo},caption={Algorithm to Get all Posts Likes from Friends}]
Double likeCount = posts.parallelstream().filter(account_id).collect((post) -> friends.getPosts(post).getLikes()).reduce()
\end{lstlisting}
While there is only one call to post at ".parallelstream()" there are N calls, N = number of likes, to friends. However, since position and assets are two different services, N number of inter-service calls are made. Each Inter-service calls travel through 2 layers, as Fig.~\ref{signalpath} shows. This then becomes N×(2L) total runtime. 
To handle It is best if the number of calls can be inter-service call can be simplified from N to 1. Hence a new solution is developed in Snippet \ref{newalgo}: 
\begin{lstlisting}[label={newalgo},caption={Optimized Algorithm}]
ConcurrentHashSet<Post> set = posts.parallelstream()
    .filter(account_id).collect((posts)) 
Double likeCount = friends.getAllPosts(set).getLikes().parrallelstream().reduce()
\end{lstlisting}
Here although 2 streams are used the application run time increased from N to 2N, but since there is only one inter service call. The total runtime becomes 2N + L. Later algorithm is used to facilitate all aggregated services. This ensure a maximum of 2×N + 2×L = 2(N+L) run time anywhere in the network, where N = number microservices of concern L = layers between these microservices.

\subsection{Memory Optimization}
\begin{figure}[htbp]
    \centerline{\includegraphics[width=\linewidth,scale=0.5]{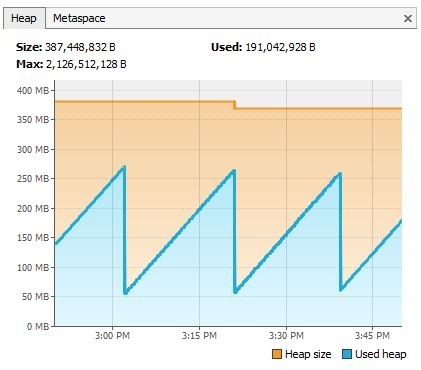}}
    \caption{Heap Usage After Optimization}
    \label{heapopt}
\end{figure}

Upon finishing the code, memory optimization becomes the next issue. While duplicating the application's java dependencies in each microservice' container makes the application only slightly smaller than the original application. For example, if the original service costs 800 MiB, the four atomic services each costs 400 MiB, and 1 BFF cost 600. Summing the size of all microservices: 4*400 + 600 = 3200mb, new microservices use four times as many memory spaces as the original application. When monitoring the size of the microservice application, we can use the guideline (6)
\begin{equation}
     (\sum_{i=1}^{n}x) >X, x_{i}<X,
\end{equation}, where x is the size of each microservice, n is the number of services, and X is the size of the original service. In short, this formula requires each microservice to be smaller the size of the monolithic application, but the total size will always be greater than the original application. The repetitive load balancing is unavoidable in containers. Since load balancing has been handled by Kubernetes' load balancer, the developers are safe to remove all load balancing mechanisms within the java app that handles loading as Kubernetes will handle the loading. First, remove the JPA library used to connect the database from the BFF because it does not connect to the database. Very quickly, the size decreased by 200 MiB. Next, the team looks at minimizing the memory size; two equations are used to model total memory use (7).

\begin{equation}
\begin{array}{l@{{}={}}l}
    memory & heap + non-heap \\ non-heap & threads \times stack + classes \times 7 \div 1000  
\end{array}
\end{equation}
By referring to the Spring website, the team finds Table \ref{springjarusage}.
\begin{table}[htbp]
    \centering
    \begin{tabular}{|c|c|c|c|c|}
    \hline
    APPLICATION & HEAP & NON HEAP & THREADS & CLASSES \\
    \hline
    Vanilla    & 22 &    50 &    25 &    6200 \\
    \hline
    Plain Java &    6 &    14 &    11 &    1500 \\
    \hline
    \end{tabular}
    \caption{Spring and Java Memory Usage}
    \label{springjarusage}
\end{table}
The developer chooses to restrict the threads to 3 and let Kubernetes expands when it needs more pods. Next, since JPA has been taken out 5000 classes are removed. As the results in Fig.~\ref{heapopt} show, eventually the memory drops to 375mbs with total heap usage cycles from 50 to 300.

\subsection{Deployment with Kuberenetes}
\begin{figure}[htbp]
    \centerline{\includegraphics[width=\linewidth,scale=0.5]{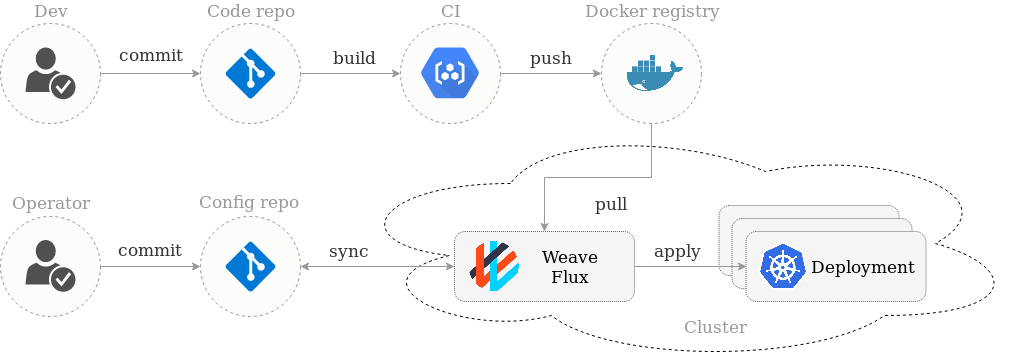}}
    \caption{Continous Integration Pipeline with Kubernetes and GitOps}
    \label{gitops}
\end{figure}

The last step is to package the deployment into Kubernetes files and figure out a way to manage the configuration. Spring offers a git config server that supports remote configuration. Put Snippet \ref{appconfig} in the application.yml file \cite{b12}: 

\begin{lstlisting}[label={appconfig}, caption={Application Level Environment Variables defined in "application.yml"}]
spring:
  cloud:
    config:
      server:
        git:
          uri: https://github.com/spring-cloud-samples/config-repo
          repos:
            development:
              pattern:
                - '*/development'
                - '*/staging'
              uri: https://github.com/development/config-repo
            staging:
              pattern:
                - '*/qa'
                - '*/production'
              uri: https://github.com/staging/config-repo
\end{lstlisting}
Now the spring app will read configuration stored in a remote git repository “https://github.com/development/config-repo”. Git config allows the deployment team to make changes outside the deployment server and not have to work with the application code. In Kubernetes, the developer will use the config map to achieve the same goal. A config map is a user-defined key-value pair store in the cluster’s etcd. Config map is visible to all pods within the cluster. For example, the user can run Snippet \label{configmap} to deploy a config map that contains key and password to a database: 
\begin{lstlisting}[label={configmap}, caption={Kubernetes Pod level Environmental Variables Defined in Config Maps}]
kind: ConfigMap
name: db_info
metadata:
  database: jdcb://db.cca.com:3306
  name: db
  password: -jyc9ep2
\end{lstlisting}
Then in the deployment file, add a reference to this config map as Snippet \ref{configmapref} does.
\begin{lstlisting}[label={configmapref}, caption={Adding Reference to Config Map on Deployment}]
apiVersion: v1
kind: deployment
metadata:
  name: account
spec:
  containers:
    - name: test
      image: k8s.gcr.io/busybox
      env:
        - name: SPECIAL_LEVEL_KEY
          valueFrom:
            configMapKeyRef: db_info
  restartPolicy: Never
\end{lstlisting}
Now in the application's "application.yaml", one can directly use values from the config map as a system environment variable. If multiple config maps are used with the same values, simply add the config map name prepended to the variable name in Snippet \ref{decoder}. 
\begin{lstlisting}[caption={Dereferencing Environmental Variables}, label={decoder}]
${name}
${db_info1.password}
${db_info2.password}
\end{lstlisting}
Configuration can be easily managed outside of the deployment. Similar to the idea of the rollout, every time the configuration is edited, the deployer will run a restart command to Kubernetes, which will bring up a new pod with the new configuration meanwhile bring one old pod down until all old pods are replaced by the new ones. Towards the conclusion of the prototype construction, an approach was discovered to build a pipeline (Fig.~\ref{gitops}) to connect the remote config repository with the Kubernetes cluster such that whenever the remote config repository is changed, no human involvement is needed to trigger the rollout. The pipeline will recognize a new change and trigger a pod restart command to Kubernetes' Istio framework. One such plugin to perform the task is Weave Flux\cite{b13}. The developers may choose to build their own pipeline from scratch. 

\section{Conclusion}
In this paper, we introduced a process to transform an existing application to a microservice system that is scalable, portable, and continually deployable. We also discussed service to service communication, data strategy, and Canary development in the new set up. Eventually, we constructed a prototype in Java using a microservice strategy. It involves adding Istio, a mesh service platform for Kubernetes. In Istio, service communicates by calling each's IP by name, when a pod exceeds its predetermined threshold, Istio automatically scale up according to the rule specified by the user. This rule lets developers filter incoming requests and set the percentage of traffic going into different versions of pods, a useful feature in the A/B split test. Although the team wishes to use Mesh expansion to incorporate new data schema into the service in the future, the team decided to settle with using service entry to open a channel for data operation. During the application, a couple of side problems was addressed. First, a methodology of minimizing inter-service calls by combining all information needed into one call, enabling a 2(N + L) runtime, where N = number of microservices, and L = number of layers between services. A quick tweak was performed to remove the unneeded library in spring to minimize jar memory usage during deployment. Lastly, the team decided to use Config map a Kubernetes object to supply needed configuration data, thus allowing the deployment team to edit configuration from a remote repository quickly.

\subsection{Recommendation}
As mentioned in 7.3, we have not had a chance to build the pipeline to enable automatic configuration rollover. Having such a feature would benefit the fluidity amongst the deployment team greatly. When two developers working on the same microservice changed a configuration at the same time, some data from the microservice may be lost due to concurrency. If A pipeline is implemented, it will establish a queue to roll over one set of changes before rolling over the next set and keep a log of all changes. When the developer team wants to roll back, a quick reversal action can be executed from the pipeline to check out the git history of the old configurations manually.

\end{document}